\documentstyle[aps,preprint,floats,epsf,epsfig]{revtex}
\begin{document}
\def\be{\begin{eqnarray}}
\def\en{\end{eqnarray}}
\def\non{\nonumber}
\def\la{\langle}
\def\ra{\rangle}
\def\nc{N_c^{\rm eff}}
\def\vp{\varepsilon}
\def\drho{\bar\rho}
\def\deta{\bar\eta}
\def\vma{{_{V-A}}}
\def\vpa{{_{V+A}}}
\def\J{{J/\psi}}
\def\B{{\cal B}}
\def\T{{\cal T}}
\def\C{{\cal C}}
\def\A{{\cal A}}
\def\E{{\cal E}}
\def\D{{D^*}}
\def\ov{\overline}
\def\Lqcd{{\Lambda_{\rm QCD}}}
\def\pr{{\sl Phys. Rev.}~}
\def\prl{{\sl Phys. Rev. Lett.}~}
\def\pl{{\sl Phys. Lett.}~}
\def\np{{\sl Nucl. Phys.}~}
\def\zp{{\sl Z. Phys.}~}
\def\lsim{ {\ \lower-1.2pt\vbox{\hbox{\rlap{$<$}\lower5pt\vbox{\hbox{$\sim$}
}}}\ } }
\def\gsim{ {\ \lower-1.2pt\vbox{\hbox{\rlap{$>$}\lower5pt\vbox{\hbox{$\sim$}
}}}\ } }

\font\el=cmbx10 scaled \magstep2{\obeylines\hfill February, 2002}

\vskip 1.5 cm

\centerline{\large\bf Implications of Recent $\ov B^0\to D^{(*)0}
X^0$ Measurements}
\bigskip
\centerline{\bf Hai-Yang Cheng}
\medskip
\centerline{Institute of Physics, Academia Sinica}
\centerline{Taipei, Taiwan 115, Republic of China}
\medskip
\centerline{and}
\medskip
\centerline{Physics Department, Brookhaven National Laboratory}
\centerline{Upton, New York 11973}
\medskip

\bigskip
\bigskip
\centerline{\bf Abstract}
\bigskip
{\small The recent measurements of the color-suppressed modes $\ov
B^0\to D^{(*)0}\pi^0$ imply non-vanishing relative final-state
interaction (FSI) phases among various $\ov B\to D\pi$ decay
amplitudes. Depending on whether or not FSIs are implemented in
the topological quark-diagram amplitudes, two solutions for the
parameters $a_1$ and $a_2$ are extracted from data using various
form-factor models. It is found that $|a_2(D\pi)|\sim 0.35-0.60$
and $|a_2(D^*\pi)|\sim 0.25-0.50$ with a relative phase of order
$60^\circ$ between $a_1$ and $a_2$. If FSIs are not included in
quark-diagram amplitudes from the outset, $a_2^{\rm eff}/a_1^{\rm
eff}$ and $a_2^{\rm eff}$ will become smaller. The large value of
$|a_2(D\pi)|$ compared to $|a_2^{\rm eff}(D\pi)|$ or naive
expectation implies the importance of long-distance FSI
contributions to color-suppressed internal $W$-emission via
final-state rescatterings of the color-allowed tree amplitude.

}

\vskip 5 cm


\pagebreak

\section{Introduction}
For some time $B\to\J K$ and $B\to \J K^*$ remain to be the only
color-suppressed $B$ meson two-body decay modes that have been
measured experimentally. Recently, the long awaited
color-suppressed decay modes $\ov B^0\to D^{(*)0}\pi^0$ are
finally measured by both Belle \cite{Belle} and CLEO \cite{CLEO}
with the $D^0\pi^0$ branching ratio larger than the upper limit
previously reported \cite{CLEOa}. The channels $\ov B^0\to
D^{(*)0}\eta$ and $\ov B^0\to D^{(*)0}\omega$ are also observed by
Belle \cite{Belle}. We shall see below that the theoretical
predictions based on the factorization approach in general are too
small to account for the observed decay rates of color-suppressed
modes $D^{(*)0}X^0$ with $X=\pi,\eta,\omega$. This has important
implications for final-state interactions (FSIs).

Under the factorization hypothesis, the nonleptonic decay
amplitudes are approximated by the factorized hadronic matrix
elements multiplied by some universal, process-independent
effective coefficients $a_i$. Based on the factorization
assumption, one can catalog the decay processes into three
classes. For class-I decays, the decay amplitudes, dominated by
the color-allowed external $W$-emission, are proportional to
$a_1\la O_1\ra_{\rm fact}$ where $O_1$ is a charged
current--charged current 4-quark operator. For class-II decays,
the decay amplitudes, governed by the color-suppressed internal
$W$-emission, are described by $a_2\la O_2\ra_{\rm fact}$ with
$O_2$ being a neutral current--neutral current 4-quark operator.
The decay amplitudes of the class-III decays involve a linear
combination of $a_1\la O_1\ra_{\rm fact}$ and $a_2\la O_2\ra_{\rm
fact}$. If factorization works, the effective coefficients $a_i$
in nonleptonic $B$ or $D$ decays should be channel by channel
independent.

What is the relation between the coefficients $a_i$ and the Wilson
coefficients in the effective Hamiltonian approach ? Under the
naive factorization hypothesis, one has
 \be \label{nf}
a_1(\mu)=c_1(\mu)+{1\over N_c}c_2(\mu), \qquad \quad
a_2(\mu)=c_2(\mu)+{1\over N_c}c_1(\mu),
 \en
for decay amplitudes induced by current-current operators
$O_{1,2}(\mu)$, where $c_{1,2}(\mu)$ are the corresponding Wilson
coefficients and $N_c$ is the number of colors. In the absence of
QCD corrections, $c_1=1$ and $c_2=0$, and hence class-II modes
governed by $a_2=1/N_c$ are obviously ``color-suppressed".
However, this naive factorization approach encounters two
principal difficulties: (i) the coefficients $a_i$ given by Eq.
(\ref{nf}) are renormalization scale and $\gamma_5$-scheme
dependent, and (ii) it fails to describe the color-suppressed
class-II decay modes. For example, the ratio $R=\Gamma(D^0\to\ov
K^0\pi^0)/\Gamma(D^0\to K^-\pi^+)$ is predicted to be only of
order $3\times 10^{-4}$ due to the smallness of $a_2$ in the naive
factorization approach, while experimentally it is measured to be
$0.55\pm 0.06$ \cite{PDG}. It is known that the decay $D^0\to\ov
K^0\pi^0$ is enhanced by two mechanisms. First, $a_2$ receives a
large nonfactorizable correction. Second, the weak decay $D^0\to
K^-\pi^+$ followed by the inelastic rescattering $K^-\pi^+\to\ov
K^0\pi^0$ can raise $\B(D^0\to\ov K^0\pi^0)$ dramatically by
lowering $\B(D^0\to K^-\pi^+)$.

Beyond naive factorization the parameters $a_{1,2}$ have the
general expression
 \be
 a_{1,2}= c_{2,1}(\mu)+{c_{1,2}(\mu)\over N_c}~
 +{\rm nonfactorizable~corrections},
 \en
where nonfactorizable corrections include vertex corrections, hard
spectator interactions involving the spectator quark of the heavy
meson, and  FSI effects from inelastic rescattering, resonance
effects, $\cdots$, etc. In the generalized factorization approach
of \cite{Ali,CT98}, one includes the vertex corrections which will
compensate the renormalization scale and $\gamma_5$-scheme
dependence of the Wilson coefficients to render $a_{1,2}$ scale
and scheme independent. Contrary to the naive one, the improved
generalized factorization scheme assumes that nonfactorizable
effects are incorporated in a process independent form. Since not
all nonfactorizable effects are calculable by perturbative QCD,
one will treat $a_1$ and $a_2$ as free parameters in the
generalized factorization approach and extract them from
experiment. The phenomenological analysis of two-body decay data
of $D$ and $B$ mesons will tell us if the generalized
factorization hypothesis  works reasonably well by studying the
variation of the parameters $a_{1,2}$ from channel to channel.

The experimental measurement of $B\to\J K$ leads to $|a_2(\J
K)|=0.26\pm 0.02$ \cite{a1a2}. This seems to be also supported by
the study of $B\to D\pi$ decays: Assuming no relative phase
between $a_1$ and $a_2$, the result $a_2\sim {\cal O}(0.20-0.30)$
\cite{a1a2,ns} is inferred from the data of $\ov B^0\to
D^{(*)+}\pi^-$ and $B^-\to D^{(*)0}\pi^-$. However, as we shall
show below, the above value of $a_2$ leads to too small decay
rates for $\ov B^0\to D^{(*)0}\pi^0$ when compared to recent
measurements. In order to account for the observation, one needs a
larger $a_2(D\pi)$ with a non-trivial phase relative to $a_1$. The
importance of FSIs has long been realized in charm decay since
some resonances are known to exist at energies close to the mass
of the charmed meson. We shall see in this work that, just as
$D^0\to\bar K^0\pi^0$, both nonfactorizable effects and FSIs are
also needed to explain the data of $\ov B^0\to D^{(*)0}\pi^0$,
though these two effects in $B$ decays are naively expected to be
not as dramatic as in the charm case.

The color-suppressed mode is a very suitable place for studying
the effect of FSIs (especially the soft one) in weak decays. The
ratio of the color-suppressed decay amplitudes with and without
FSIs is $R_{\ov K\pi} \equiv|A(D^0\to \ov K^0\pi^0)/A(D^0\to \ov
K^0\pi^0)_{\rm without~FSIs}|\approx 2.0$ and the relative phase
between $D^0\to \ov K^0\pi^0$ and $D^0\to K^-\pi^+$ is about
$150^\circ$. It is expected that for $\ov B\to D\pi$ decay,
$R_{D\pi}$ and the relative phase among decay amplitudes will
become smaller. The recent measurement of the $\ov B^0\to
D^0\pi^0$ mode allows us to determine the above two quantities. We
shall see that although the relative phase among $\ov B\to D\pi$
decay amplitudes becomes smaller, $R_{D^{(*)}\pi}$ does not
decrease in a significant way from charm to bottom case. The
implications and related physics will be discussed below in
details.

\section{Factorization}
We begin with by considering the branching ratios of the
color-suppressed modes $\ov B^0\to D^{(*)0}X^0$
$(X=\pi,\eta,\omega)$ within the framework of the factorization
approach. The $\ov B^0\to D^0\pi^0$ amplitude is given by
 \be \label{D0pi0amp}
 A(\ov B^0\to D^0\pi^0) &=& {1\over\sqrt{2}}(-\C+\E),
 \en
where $\C,~\E$ are color-suppressed internal $W$-emission and
$W$-exchange amplitudes, respectively. In terms of the factorized
hadronic matrix elements, they read
 \be \label{CE}
{\cal C} &=& i
{G_F\over\sqrt{2}}\,V_{cb}V_{ud}^*\,a_2(D\pi)(m_B^2-m_\pi^2)f_D
F_0^{B\pi}(m_D^2), \non \\ {\cal E} &=& i
{G_F\over\sqrt{2}}\,V_{cb}V_{ud}^*\,a_2(D\pi)(m_D^2-m_\pi^2)f_B
F_0^{0\to D\pi}(m_B^2),
 \en
where $a_2(D\pi)$ is a parameter to be determined from experiment.
The annihilation form factor $F_0^{0\to D\pi}(m_B^2)$ is expected
to be suppressed at large momentum transfer, $q^2=m_B^2$,
corresponding to the conventional helicity suppression. Based on
the argument of helicity and color suppression, one may therefore
neglect short-distance (hard) $W$-exchange contributions. However,
it is not clear if the long-distance contribution to $W$-exchange
is also negligible. Likewise,
 \be
 A(\ov B^0\to D^0\eta) &=& i{G_F\over\sqrt{2}}\,V_{cb}V_{ud}^*\,a_2(D\eta)(m_B^2-m_\eta^2)f_D
F_0^{B\eta}(m_D^2), \non \\
 A(\ov B^0\to D^{*0}\pi^0) &=& -{G_F\over\sqrt{2}}\,V_{cb}V_{ud}^*\,a_2(D^*\pi)\sqrt{2}m_{D^*}f_{D^*}
F_1^{B\pi}(m_{D^*}^2), \\
 A(\ov B^0\to D^0\omega) &=& {G_F\over\sqrt{2}}\,V_{cb}V_{ud}^*\,a_2(D\omega)2m_\omega
 f_D A_0^{B\omega}(m_D^2),  \non
 \en
 and
 \be
 A(\ov B^0\to D^{*0}\omega) &=& -i{G_F\over\sqrt{2}}\,V_{cb}V_{ud}^*a_2
 (D^*\omega)f_\D m_\D\Bigg[ (\vp^*_\D\cdot\vp^*_\omega)
(m_{B}+m_{\omega})A_1^{ B\omega}(m_{\D}^2)  \non \\
&-& (\vp^*_\D\cdot p_{_{B}})(\vp^*_\omega \cdot p_{_{B}}){2A_2^{
B\omega}(m_{\D}^2)\over m_{B}+m_{\omega} } +
i\epsilon_{\mu\nu\alpha\beta}\vp^{*\mu}_\omega\vp^{*\nu}_\D
p^\alpha_{_{B}} p^\beta_1\,{2V^{ B\omega}(m_{\D}^2)\over
m_{B}+m_\omega }\Bigg].
 \en
Here factorization implies a universal $a_2$, namely,
$a_2(D^*\omega)=a_2(D\omega)=a_2(D\eta)=a_2(D^*\pi)=a_2(D\pi)$. In
naive factorization, $a_2$ is not only small, of order 0.10, but
also renormalization scale and scheme dependent. In the
generalized factorization approach, the scale- and
scheme-independent $a_2$ can be extracted from experiment and the
factorization hypothesis is tested by studying $a_2$ to see if it
is process independent or insensitive.

To proceed, we shall consider four distinct form-factor models:
the Neubert-Rieckert-Stech-Xu (NRSX) model \cite{NRSX}, the
relativistic light-front (LF) quark model \cite{cch}, the
Neubert-Stech (NS) model \cite{ns}, and the Melikhov-Stech (MS)
model based on the constituent quark picture \cite{Melikhov}. The
NRSX model takes the Bauer-Stech-Wirbel (BSW) model \cite{bsw}
results for the form factors at zero momentum transfer but makes a
different ansatz for their $q^2$ dependence, namely, a dipole
behavior is assumed for the form factors $F_1,~A_0,~A_2,~V$,
motivated by heavy quark symmetry, and a monopole dependence for
$F_0,A_1$, where we have followed the definition of form factors
given in \cite{bsw}. For reader's convenience, the values of
relevant form factors are listed in Table I (see \cite{a1a2} for
some details about the NS model).

The form factors for $B\to\eta$ and $B\to\eta'$ transitions have
been calculated by BSW \cite{bsw} in a relativistic quark model.
However, in their relativistic quark model calculation of $B\to
\eta^{(')}$ transitions, BSW considered only the $u\bar u$
component of the $\eta$ and $\eta'$; that is, the form factors
calculated by BSW are actually $F_0^{B\eta_{u\bar u}}$ and
$F_0^{B\eta'_{u\bar u}}$ induced from the $b\to u$ transition. It
is thus more natural to consider the flavor basis of $\eta_q$ and
$\eta_s$ defined by
 \be
 \eta_q={1\over\sqrt{2}}(u\bar u+d\bar d),\qquad\quad
 \eta_s=s\bar s.
 \en
The wave functions of the $\eta$ and $\eta'$ are given by
 \be
 \left(\matrix{ \eta \cr \eta'\cr}\right)=\left(\matrix{ \cos\phi & -\sin\phi \cr
 \sin\phi & \cos\phi\cr}\right)\left(\matrix{\eta_q \cr \eta_s
 \cr}\right),
 \en
where $\phi=\theta+{\rm arctan}\sqrt{2}$, and $\theta$ is the
$\eta\!-\!\eta'$ mixing angle in the octet-singlet basis. The
physical form factors then have the simple expressions:
 \be \label{Beta}
F_{0,1}^{B\eta}={1\over\sqrt{2}}\cos\phi \,F_{0,1}^{B\eta_{u\bar
u}}, \qquad && F_{0,1}^{B\eta'}={1\over\sqrt{2}}\sin\phi\,
F_{0,1}^{B\eta'_{u\bar u}}.
 \en
Using $F_0^{B\eta_{u\bar u}}(0)=0.307$ and $F_0^{B\eta'_{u\bar
u}}(0)=0.254$ obtained from \cite{bsw} and the mixing angle
$\phi=39.3^\circ$ (or $\theta=-15.4^\circ$) \cite{Kroll} we find
$F_0^{B\eta}(0)=0.168$ and $F_0^{B\eta'}(0)=0.114$ in the BSW
model and hence the NRSX model. For other form-factor
models,\footnote{The form factors $F_0^{B\eta}(m_D^2)=0.28$ and
$F_1^{B\eta}(m_\D^2)=0.33$ for the NS model obtained in
\cite{Deandrea} are larger than ours by about a factor of 2.} we
shall apply the relation based on isospin-quartet symmetry
 \be \label{isospinquart}
 F_{0,1}^{B\eta_{u\bar u}}=F_{0,1}^{B\to\eta'_{u\bar
 u}}=F_{0,1}^{B\pi}
 \en
and Eq. (\ref{Beta}) to obtain the physical $B-\eta$ and $B-\eta'$
transition form factors.

{\squeezetable
\begin{table}[ht]
\caption{Form factors in various form-factor models. Except for
the NRSX model, the relations
$A_i^{B\omega}(q^2)=A_i^{B\rho^0}(q^2)$ $(i=0,1,2)$ and
$V^{B\omega}(q^2)=V^{B\rho^0}(q^2)$ are assumed in all the
form-factor models. The pion in the $B-\pi$ transition is referred
to the charged one.}
 \footnotesize
\begin{center}
\begin{tabular}{l c c c c c c c c c c c}
& $F_0^{B\pi}(m_D^2)$ & $F_1^{B\pi}(m^2_{D^*})$ &
$F_0^{B\eta}(m_D^2)$ & $F_1^{B\eta}(m_\D^2)$ & $F_0^{BD}(m_\pi^2)$
 & $A_0^{B\D}(m_\pi^2)$ &
$A_0^{B\omega}(m_D^2)$ & $A_1^{B\omega}(m_\D^2)$ &
$A_2^{B\omega}(m_\D^2)$ & $V^{B\omega}(m_\D^2)$ \\ \hline
 NRSX & 0.37 & 0.45 & 0.19 & 0.23 & 0.69  & 0.62 & 0.26 & 0.23 & 0.27 & 0.32 \\
 LF   & 0.34 & 0.39 & 0.18 & 0.22 & 0.70  & 0.73 & 0.25 & 0.16 & 0.17 & 0.28 \\
 MS   & 0.32 & 0.36 & 0.17 & 0.20 & 0.67  & 0.69 & 0.26 & 0.20 & 0.21 & 0.28 \\
 NS   & 0.27 & 0.32 & 0.15 & 0.18 & 0.63  & 0.64 & 0.22 & 0.20 & 0.22 & 0.22 \\
\end{tabular}
\end{center}
\end{table}
}

As mentioned in the Introduction, in the absence of a relative
phase between $a_1$ and $a_2$, a value of $a_2$ in the range of
0.20 to 0.30 is inferred from the data of $\ov B^0\to
D^{(*)+}\pi^-$ and $B^-\to D^{(*)0}\pi^-$. For definiteness, we
shall use the representative value $a_2=0.25$ for the purpose of
illustration. The calculated branching ratios for $\ov B^0\to
D^{(*)0}X^0$ are shown in Table II for $f_D=200$ MeV and
$f_\D=230$ MeV. Evidently, the predicted rates for color
suppressed modes are too small compared to recent measurements. It
should be stressed that if there is no relative phase between
$a_1$ and $a_2$, then one cannot increase $a_2$ arbitrarily to fit
the data as this will enhance the decay rate of the $\Delta I=3/2$
mode $B^-\to D^{(*)0}\pi^-$ and destroy the agreement between
theory and experiment for the charged mode. For example, fitting
$a_2$ to the data of $D^0\pi^0$ without FSIs will yield $a_2=0.45$
in the MS model, which in turn implies $\B(B^-\to
D^0\pi^-)=7.9\times 10^{-3}$ and this is obviously too large
compared to the experimental value $(5.3\pm0.5)\times 10^{-3}$
\cite{PDG}. In this case, one needs FSIs to convert $D^+\pi^-$
into $D^0\pi^0$. In contrast, if $a_2$ is of order 0.45, then a
relative strong phase between $a_1$ and $a_2$ will be needed in
order not to over-estimate the $D^0\pi^-$ rate. In either case, we
conclude that FSIs are the necessary ingredients for understanding
the data.

\begin{table}[ht]
\caption{Predicted branching ratios (in units of $10^{-4}$) of
$\ov B^0\to D^{(*)0}X^0$ $(X=\pi,\eta,\omega)$ in the generalized
approach with various form-factor models for $a_2=0.25$, $f_D=200$
MeV and $f_\D=230$ MeV.
 }
\begin{center}
\begin{tabular}{l l l l l  l c }
&  &  &  &  & \multicolumn{2}{c}{Experiments}   \\ \cline{6-7}
\raisebox{1.5ex}[0cm][0cm]{Decay mode} &
\raisebox{1.5ex}[0cm][0cm]{NRSX} & \raisebox{1.5ex}[0cm][0cm]{LF}
& \raisebox{1.5ex}[0cm][0cm]{MS} & \raisebox{1.5ex}[0cm][0cm]{NS}
& Belle \cite{Belle} & CLEO \cite{CLEO} \\ \hline
 $\ov B^0\to D^0\pi^0$ & 1.13& 0.93 & 0.82 & 0.58  & $3.1\pm0.4\pm0.5$ &
 $2.74^{+0.36}_{-0.32}\pm 0.55$ \\
 $\ov B^0\to D^{*0}\pi^0$ & 1.57 & 1.20 & 1.01 & 0.80 & $2.7^{+0.8+0.5}_{-0.7-0.6}$ &
 $2.20^{+0.59}_{-0.52}\pm0.79$ \\
 $\ov B^0\to D^0\eta$ & 0.55 & 0.53 & 0.48 & 0.34 & $1.4^{+0.5}_{-0.4}\pm 0.3$ & \\
 $\ov B^0\to D^{*0}\eta$ & 0.76 & 0.68 & 0.58 & 0.46 & $2.0^{+0.9}_{-0.8}\pm0.4$ & \\
 $\ov B^0\to D^0\omega$ & 0.76 & 0.71 & 0.76 & 0.54 & $1.8\pm0.5^{+0.4}_{-0.3}$ & \\
 $\ov B^0\to D^{*0}\omega$ & 1.60 & 1.16& 1.75& 1.35 & $3.1^{+1.3}_{-1.1}\pm0.8$ &
 \\
\end{tabular}
\end{center}
\end{table}

\section{Extraction of $\lowercase{a_1}$ and $\lowercase{a_2}$}
In this section we will extract the parameters $a_1$ and $a_2$ in
two different approaches. In the first approach, the topological
amplitudes are assumed to incorporate all the information of
strong interactions. Therefore, $a_{1,2}$ thus determined already
include the effects of FSIs. In the second approach, one will
assume that quark-diagram topologies in their original forms do
not include FSIs from the outset.

\subsection {Direct analysis}
In terms of the quark-diagram topologies $\T$, $\C$ and $\E$,
where $\T$ is the color-allowed external $W$-emission amplitude,
the other $\ov B\to D\pi$ amplitudes can be expressed as
 \be \label{Dpiamp}
 A(\ov B^0\to D^+\pi^-) &=& \T+\E,\non\\
 A(B^-\to D^0\pi^-) &=& \T+\C,
 \en
and they satisfy the isospin triangle relation
 \be
 A(\ov B^0\to D^+\pi^-)=\sqrt{2}A(\ov B^0\to D^0\pi^0)+A(B^-\to D^0\pi^-).
 \en
In writing Eqs. (\ref{D0pi0amp}) and (\ref{Dpiamp}) it has been
assumed that the topologies $\T,~\C,~\E$ include the information
of all strong interactions for physical $\ov B\to D\pi$ amplitudes
(for an earlier discussion of quark-diagram amplitudes, see
\cite{CC}). Now since all three sides of the $\ov B\to D\pi$
triangle are measured, we are able to determine the relative
phases among the decay amplitudes. Using the data \cite{PDG}
 \be
\B(\ov B^0\to D^+\pi^-)=(3.0\pm0.4)\times 10^{-3}, && \qquad
\B(B^-\to D^0\pi^-)=(5.3\pm 0.5)\times 10^{-3}, \non \\
 \B(\ov B^0\to D^{*+}\pi^-)=(2.76\pm0.21)\times 10^{-3}, &&
\qquad \B(B^-\to D^{*0}\pi^-)=(4.6\pm 0.4)\times 10^{-3},
\label{data}
 \en
and the combined value of Belle and CLEO for the neutral modes
(see Table II)
 \be
\B(\ov B^0\to D^0\pi^0)=(2.92\pm 0.46)\times 10^{-4}, \qquad\quad
\B(\ov B^0\to D^{*0}\pi^0)=(2.47\pm 0.67)\times 10^{-4},
 \en
we find (only the central values for phase angles are shown here)
 \be \label{CET}
 \left.{\C-\E\over \T+\E}\right|_{D\pi}=(0.44\pm0.05)\, e^{i59^\circ}, && \qquad\quad
  \left.{\C-\E\over \T+\C}\right|_{D\pi}=(0.34\pm0.03)\, e^{i37^\circ},
  \non \\
 \left.{\C-\E\over \T+\E}\right|_{D^*\pi}=(0.42\pm0.06)\, e^{i63^\circ},
 &&\qquad\quad \left.{\C-\E\over \T+\C}\right|_{D^*\pi}=(0.34\pm0.05)\, e^{i44^\circ},
 \en
where we have employed the $B$ meson lifetimes given in
\cite{PDG}.

The same phases also can be obtained from the isospin analysis.
Decomposing the physical amplitudes into their isospin amplitudes
yields
 \be \label{isospin} A(\ov B^0\to D^+\pi^-) &=& \sqrt{2\over
3}A_{1/2}+\sqrt{1\over 3}A_{3/2},   \non  \\
A(\ov B^0\to D^0\pi^0) &=& \sqrt{1\over 3}A_{1/2} -\sqrt{2\over
3}A_{3/2}, \\ A(B^-\to D^0\pi^-) &=& \sqrt{3}A_{3/2}. \non
 \en
The isospin amplitudes are related to the topological
quark-diagram amplitudes via
 \be \label{isospinrel}
 A_{1/2}={1\over\sqrt{6}}(2\T-\C+3\E), \qquad\qquad
 A_{3/2}={1\over \sqrt{3}}(\T+\C).
 \en
Intuitively, the phase shift difference between $A_{1/2}$ and
$A_{3/2}$, which is of order $90^\circ$ for $D\to \ov K\pi$ modes
(see below), is expected to play a minor role in the energetic
$B\to D\pi$ decay, the counterpart of $D\to\ov K\pi$ in the $B$
system, as the decay particles are moving fast, not allowing
adequate time for final-state interactions. Applying the relations
(see e.g. \cite{ns})
 \be
 |A_{1/2}|^2&=& |A(\ov B^0\to D^+\pi^-)|^2+|A(\ov B^0\to
 D^0\pi^0)|^2-{1\over 3}|A(B^-\to D^0\pi^-)|^2, \non \\
 |A_{3/2}|^2&=&{1\over 3}|A(B^-\to D^0\pi^-)|^2, \\
 \cos(\delta_{1/2}-\delta_{3/2}) &=& {3|A(\ov B^0\to
 D^+\pi^-)|^2-2|A_{1/2}|^2-|A_{3/2}|^2\over 2\sqrt{2}
 |A_{1/2}||A_{3/2}|}, \non
 \en
we obtain
 \be \label{A13}
 \left.{A_{1/2}\over \sqrt{2}A_{3/2}}\right|_{D\pi}=(0.70\pm0.10)\,e^{i29^\circ},
 \qquad\quad
 \left.{A_{1/2}\over \sqrt{2}
 A_{3/2}}\right|_{D^*\pi}=(0.74\pm0.07)\,e^{i29^\circ}.
 \en
Similar results are also obtained before in \cite{Xing,NP} using
the preliminary Belle and CLEO measurements. It is easy to check
that the ratio $(\C-\E)/(\C+\E)$ in Eq. (\ref{CET}) follows from
Eqs. (\ref{isospinrel}) and (\ref{A13}). It is also interesting to
compare the above results with that for $D\to \ov K^{(*)}\pi$
decays \cite{PDG}:
 \be
 \left.{A_{1/2}\over \sqrt{2}A_{3/2}}\right|_{\ov K\pi}= (2.70\pm0.14)\,e^{i90^\circ},
 \qquad\quad
 \left.{A_{1/2}\over \sqrt{2}A_{3/2}}\right|_{\ov K^*\pi}=
 (3.97\pm0.25)\,e^{i104^\circ}.
 \en
The smaller isospin phase shift difference in $B$ decays is in
accord with expectation. Notice that while $\Delta I=1/2$ and 3/2
amplitudes in $\ov B\to D^{(*)}\pi$ are of the same size, the
$D\to\ov K\pi$ decays are dominated by the isospin $\Delta I=1/2$
amplitude. In the heavy quark limit, the ratio of
$A_{1/2}/(\sqrt{2}A_{3/2})$ approaches to unity \cite{NP}.
Evidently, the charm system exhibits a more deviation than the $B$
system from the heavy quark limit, as expected.

The ratio of $a_2/a_1$ can be extracted from Eq. (\ref{CET}) or
Eq. (\ref{A13}). Noting that the factorized color-allowed tree
amplitude reads
 \be \label{T}
 \T= i{G_F\over\sqrt{2}}\,V_{cb}V_{ud}^*\,a_1(D\pi)(m_B^2-m_D^2)f_\pi
F_0^{BD}(m_\pi^2),
 \en
and neglecting $W$-exchange contributions, we get
 \be \label{a21Dpi}
 \left.{a_2\over a_1}\right|_{D\pi}&=& (0.44\pm0.05)\,e^{i59^\circ}\times
 {f_\pi\over f_D}\,{m_B^2-m_D^2\over
 m_B^2-m_\pi^2}\,{F_0^{BD}(m_\pi^2)\over
 F_0^{B\pi}(m_D^2)} \non \\
&=& {1-\left.{A_{1/2}\over \sqrt{2}A_{3/2}}\right|_{D\pi}\over
 {1\over 2}+\left.{A_{1/2}\over \sqrt{2}A_{3/2}}\right|_{D\pi} }\,{f_\pi\over f_D}\,{m_B^2-m_D^2\over
 m_B^2-m_\pi^2}\,{F_0^{BD}(m_\pi^2)\over
 F_0^{B\pi}(m_D^2)}.
 \en
Likewise, for the $\ov B\to D^*\pi$ decays
 \be \label{a21D*pi}
 \left.{a_2\over a_1}\right|_{D^*\pi}&=& (0.42\pm0.06)\,e^{i63^\circ}\times
 {f_\pi\over f_\D}\,{A_0^{BD^*}(m_\pi^2)\over
 F_1^{B\pi}(m_\D^2)}
 = {1-\left.{A_{1/2}\over \sqrt{2}A_{3/2}}\right|_{D^*\pi}\over
 {1\over 2}+\left.{A_{1/2}\over \sqrt{2}A_{3/2}}\right|_{D^*\pi} }\,{f_\pi\over
 f_\D}\,{A_0^{BD^*}(m_\pi^2)\over
 F_1^{B\pi}(m_\D^2)}.
 \en
With the form factors given in various models, we are ready to
extract $a_1$ and $a_2$ from the experimental data. The results
are shown in Table III and the parameter $a_2$ falls into the
range of $|a_2(D\pi)|\sim 0.35-0.60$ and $|a_2(D^*\pi)|\sim
0.25-0.50$. Note that the phases of $a_2/a_1$, $59^\circ$ for the
$D\pi$ system and $63^\circ$ for $D^*\pi$, are slightly different
from that given in \cite{NP} based on the preliminary Belle and
CLEO data. We see that although $|a_2(D\pi)|$ and $|a_2(D^*\pi)|$
agree to within one standard deviation, there is a tendency that
the former is slightly larger than the latter. Hence,
nonfactorizable effects could be process dependent, recalling that
the experimental value for $B\to\J K$ is $|a_2(\J K)|=0.26\pm
0.02$ \cite{a1a2}.

{\squeezetable
\begin{table}[ht]
\caption{Extraction of the parameters $a_1$ and $a_2$ from the
measured $B\to D^{(*)}\pi$ rates by assuming a negligible
$W$-exchange contribution. Note that $a_2(D\pi)$ and $a_2(D^*\pi)$
should be multiplied by a factor of (200 MeV/$f_D$) and (230
MeV/$f_\D$), respectively.
 }
  \footnotesize
\begin{center}
\begin{tabular}{l | c c c | c c c }
 Model~~ & $|a_1(D\pi)|$ & $|a_2(D\pi)|$ & $a_2(D\pi)/a_1(D\pi)$~ & $|a_1(\D\pi)|$
 & $|a_2(\D\pi)|$ & $a_2(\D\pi)/a_1(\D\pi)$ \\ \hline
 NRSX & $0.85\pm0.06$ & $0.40\pm0.05$ & $(0.47\pm0.05)\,{\rm exp}(i59^\circ)$
 & $0.94\pm0.04$ & $0.31\pm0.04$ & $(0.33\pm0.04)\,{\rm exp}(i63^\circ)$ \\
 LF & $0.84\pm0.06$ & $0.44\pm0.06$ & $(0.53\pm0.06)\,{\rm exp}(i59^\circ)$
 & $0.80\pm0.03$ & $0.36\pm0.05$ & $(0.45\pm0.06)\,{\rm exp}(i63^\circ)$ \\
 MS & $0.88\pm0.06$ & $0.47\pm0.06$ & $(0.53\pm0.06)\,{\rm exp}(i59^\circ)$
 & $0.85\pm0.03$ & $0.389\pm0.05$ & $(0.46\pm0.06)\,{\rm exp}(i63^\circ)$ \\
 NS & $0.93\pm0.06$ & $0.56\pm0.07$ & $(0.60\pm0.07)\,{\rm exp}(i59^\circ)$
 & $0.91\pm0.03$ & $0.44\pm0.06$ & $(0.48\pm0.06)\,{\rm exp}(i63^\circ)$ \\
\end{tabular}
\end{center}
\end{table}
}

Ideally, the parameters $a_1$ and $a_2$ will be more precisely
determined if the topologies $\T,~\C$ and $\E$ can be individually
extracted from experiment. Indeed, this is the case for charm
decays where $\T,~\C$ and $\E$ can be determined from $D\to \ov
K\pi$, $D\to \ov K\eta$ and $D\to\ov K\eta'$ decays based on SU(3)
flavor symmetry and it is found that $|\T|:|\C|:|\E|\sim 1.7: 1.3:
1.0$ \cite{Rosner}. Hence, the $W$-exchange amplitude that
receives short-distance and long-distance contributions is not
negligible at all in charm decay.\footnote{From \cite{Rosner} one
can deduce that $xa_2/a_1=\C/\T=(0.73\pm0.05)\,{\rm
exp}(i152^\circ)$ for $D\to PP$ decays without making any
assumption on $W$-exchange, to be compared with the value
$(1.05\pm0.05){\rm exp}(i149^\circ)$ obtained in \cite{NP} by
neglecting $W$-exchange.} Unfortunately, one cannot extract those
three quark-diagram amplitudes for $B$ decays since the decay
amplitudes of $\ov B^0\to D^0(\eta,\eta')$ are proportional to
$(\C+\E)$, while $D^0\pi^0$ is governed by $(-\C+\E)$ [see Eq.
(\ref{CE})]. Therefore, the quark-diagram amplitudes $\C$ and $\E$
cannot be disentangled. Nevertheless, an accurate measurement of
$D^0(\eta,\eta')$ will enable us to test the importance of
$W$-exchange in $\ov B\to D\pi$ decays.

In principle, $a_1$ can be determined in a model-independent way
from the measurement of the ratio of the decay rate of
color-allowed modes to the differential semileptonic distribution
at the appropriate $q^2$ \cite{Bjorken}:
 \be   \label{S}
 S_h^{(*)}\equiv {{\cal B}(\ov B^0\to
D^{(*)+}h^-)\over d{\cal B}(\ov B^0\to D^{(*)+}\ell^-\bar
\nu)/dq^2\Big|_{q^2=m^2_h} }=6\pi^2\,a_1^2f_h^2
|V_{ij}|^2Y_h^{(*)},
 \en
where $V_{ij}$ is the relevant CKM matrix element and the
expression of $Y_h^{(*)}$ can be found in \cite{ns}. Since the
ratio $S_h^{(*)}$ is independent of $V_{cb}$ and form factors, its
experimental measurement can be utilized to fix $a_1$ in a
model-independent manner, provided that $Y_h^{(*)}$ is also
independent of form-factor models. Based on the earlier CLEO data,
it is found that $a_1(D\pi)=0.93\pm 0.10$ and $a_1(D^*\pi)=1.09\pm
0.07$ \cite{a1a2}. Needless to say, the forthcoming measurements
from BaBar, Belle and CLEO will enable us to extract the
model-independent $a_1$ more precisely. Note that QCD
factorization predicts $a_1(D^{(*)}\pi)\approx 1.05$ in the heavy
quark limit \cite{BBNS}.

Assuming $a_2(D^{(*)}\eta^{(')})=a_2(D^{(*)}\pi)$ we see from
Table IV that the predicted branching ratios of  $\ov B^0\to
D^{(*)0}\eta$ are consistent with experiment. Note that the
predicted rates of $D^{(*)0}(\eta, \eta')$ are the same for LF, MS
and NS models since $a_2(D\pi)F_0^{B\pi}(m_D^2)$ is model
independent [see Eq. (\ref{CE})] and the form factors
$F_0^{B\eta_{0}}$ and $F_0^{B\eta_8}$ are assumed to be
proportional to $F_0^{B\pi}$ in these models.

\begin{table}[ht]
\caption{Predicted branching ratios (in units of $10^{-4}$) of
$\ov B^0\to D^{(*)0}(\eta,\eta')$  in various form-factor models
by assuming $a_2(D^{(*)}\eta^{(')})=a_2(D^{(*)}\pi)$.
 }
\begin{center}
\begin{tabular}{l c c c c  l c }
{Decay mode} & NRSX & LF & MS & NS & Experiment [1]
\\ \hline
 $\ov B^0\to D^0\eta$ & $1.43\pm0.24$ & $1.69\pm0.28$ & $1.69\pm0.28$
 & $1.69\pm0.28$ & $1.4^{+0.5}_{-0.4}\pm 0.3$  \\
 $\ov B^0\to D^{*0}\eta$ & $1.20\pm0.29$ & $1.41\pm0.35$ & $1.41\pm0.35$
 & $1.41\pm0.35$ & $2.0^{+0.9}_{-0.8}\pm0.4$  \\
 $\ov B^0\to D^0\eta'$ & $0.89\pm0.15$ & $1.05\pm0.18$ & $1.05\pm0.18$
 & $1.05\pm0.18$ &   \\
 $\ov B^0\to D^{*0}\eta'$ & $0.72\pm0.18$ & $0.85\pm0.21$ & $0.85\pm0.21$ & $0.85\pm0.21$ &
 \\
\end{tabular}
\end{center}
\end{table}

\subsection{Effective parameters $a_1^{\rm eff}$ and $a_2^{\rm
eff}$} Thus far we have assumed that quark-diagram topologies
include all strong-interaction effects including FSIs. It is
equally well to take a different point of view on the
quark-diagram topologies, namely, their original forms do not
include FSIs from the outset. In this case, there is no relative
strong phase between the isospin amplitudes $A_{1/2}$ and
$A_{3/2}$ given by Eq. (\ref{isospinrel}). Next, one puts isospin
phase shifts into Eq. (\ref{isospin}) to get
 \be \label{isospinfsi}
 A(\ov B^0\to D^+\pi^-)_{\rm FSI} &=& \sqrt{2\over
3}A_{1/2}e^{i\delta_{1/2}}+\sqrt{1\over 3}A_{3/2}e^{i\delta_{3/2}},   \non  \\
A(\ov B^0\to D^0\pi^0)_{\rm FSI} &=& \sqrt{1\over
3}A_{1/2}e^{i\delta_{1/2}} -\sqrt{2\over
3}A_{3/2}e^{i\delta_{3/2}},
\\ A(B^-\to D^0\pi^-)_{\rm FSI} &=&
\sqrt{3}A_{3/2}e^{i\delta_{3/2}}, \non
 \en
where the subscript ``FSI" indicates that the physical amplitudes
take into account the effects of FSIs. This is motivated by
comparing the experimental results with the calculated isospin
amplitudes under the factorization approximation. Neglecting
inelastic scattering, one can then extract the coefficients
$a_{1,2}^{\rm eff}$ from a comparison of the measured and
calculated isospin amplitudes \cite{ns}. It is straightforward to
show that
 \be \label{ampfsi}
 A(\ov B^0\to D^0\pi^0)_{\rm FSI} &=& A(\ov B^0\to D^0\pi^0)+{2{\cal T}- {\cal
C}+3\E\over 3\sqrt{2}}\left(
e^{i(\delta_{1/2}-\delta_{3/2})}-1\right),
\non \\
A(\ov B^0\to D^+\pi^-)_{\rm FSI} &=& A(\ov B^0\to
D^+\pi^-)+{2{\cal T}- {\cal C}+3\E\over 3}\left(
e^{i(\delta_{1/2}-\delta_{3/2})}-1\right),
 \en
where we have dropped the overall phase $e^{i\delta_{3/2}}$. The
quark-diagram amplitudes $\T,~\C,~\E$ in Eq. (\ref{ampfsi}) have
the same expressions as before except that $a_{1,2}$ in Eqs.
(\ref{CE}) and (\ref{T}) are replaced by the real parameters
$a_{1,2}^{\rm eff}$. The latter do not contain FSI effects and are
defined for $\delta_{1/2}=\delta_{3/2}=0$
\cite{Xing}.\footnote{The distinction of hard and soft FSI phases
in principle cannot be done in a systematical way. For example, a
sizable ``hard" strong-interaction phase for $a_2$ in $B\to\pi\pi$
decay is calculable in the QCD factorization approach. However,
$a_2$ is not computable for $\ov B\to D\pi$ and hence its strong
phase is most likely soft.} In other words, the parameters
$a_{1,2}^{\rm eff}$ are defined when FSIs are not imposed to the
topological quark diagram amplitudes.

The isospin phase difference in Eq. (\ref{ampfsi}) is $29^\circ$
for both $\ov B\to D\pi$ and $\ov B\to D^*\pi$. It is easily seen
that $a_2^{\rm eff}/a_1^{\rm eff}$ is determined from the second
line of Eqs. (\ref{a21Dpi}) and (\ref{a21D*pi}) but without a
phase for the ratio $A_{1/2}/(\sqrt{2}A_{3/2})$. For example,
$a_2^{\rm eff}/a_1^{\rm eff}$ for $\ov B\to D\pi$ is given by
 \be
 \left.{a_2^{\rm eff}\over a_1^{\rm eff}}\right|_{D\pi}
= {1-\left|{A_{1/2}\over \sqrt{2}A_{3/2}}\right|_{D\pi}\over
 {1\over 2}+\left|{A_{1/2}\over \sqrt{2}A_{3/2}}\right|_{D\pi}
 }\,{f_\pi\over f_D}\,{m_B^2-m_D^2\over
 m_B^2-m_\pi^2}\,{F_0^{BD}(m_\pi^2)\over
 F_0^{B\pi}(m_D^2)}.
 \en
The results are shown in Table V. Obviously $a^{\rm eff}_2/a^{\rm
eff}_1$ and $a^{\rm eff}_2$ are smaller than the previous
solution.

\begin{table}[ht]
\caption{Extraction of the parameters $a^{\rm eff}_1$ and $a^{\rm
eff}_2$ from the measured $B\to D^{(*)}\pi$ rates. Note that
$a_2^{\rm eff}(D\pi)$ and $a_2^{\rm eff}(D^*\pi)$ should be
multiplied by a factor of (200 MeV/$f_D$) and (230 MeV/$f_\D$),
respectively.
 }
\begin{center}
\begin{tabular}{l | c c c | c c c }
 Model & $a_1^{\rm eff}(D\pi)$ & $a_2^{\rm eff}(D\pi)$ & $a_2^{\rm eff}(D\pi)
 /a_1^{\rm eff}(D\pi)$ & $a_1^{\rm eff}(\D\pi)$
 & $a_2^{\rm eff}(\D\pi)$ & $a_2^{\rm eff}(\D\pi)/a_1^{\rm eff}(\D\pi)$ \\ \hline
 NRSX & $0.88\pm0.06$ & $0.23\pm0.08$ & $0.26\pm0.09$ & $0.97\pm0.04$ & $0.16\pm0.04$ & $0.17\pm0.04$ \\
 LF & $0.87\pm0.06$ & $0.25\pm0.09$ & $0.29\pm0.10$ & $0.83\pm0.03$ & $0.18\pm0.05$ & $0.22\pm0.05$ \\
 MS & $0.91\pm0.06$ & $0.27\pm0.10$ & $0.30\pm0.10$ & $0.87\pm0.03$ & $0.20\pm0.05$ & $0.23\pm0.06$ \\
 NS & $0.96\pm0.06$ & $0.32\pm0.12$ & $0.34\pm0.11$ & $0.94\pm0.04$ & $0.22\pm0.06$ & $0.24\pm0.06$ \\
\end{tabular}
\end{center}
\end{table}

\subsection {Comparison}
We are ready to compare the above two different types of
approaches. In the type-I solution, $D^{(*)0}\pi^0$ rates are
accommodated because of an enhanced $|a_2(D^{(*)0}\pi)|$. The
branching ratio of $D^{(*)0}\pi^-$ is not over-estimated owing to
a relative strong phase between $a_1$ and $a_2$. In the type-II
solution, although $a_2^{\rm eff}$ is smaller than the magnitude
of $a_2$, the $D^{(*)0}\pi^0$ states gain a feedback from
$D^{(*)+}\pi^-$ via FSIs. \footnote{Recently, it has been
suggested in \cite{CHY} that quasi-elastic scatterings of
$D^{(*)}P\to D^{(*)}P$ and $DV\to DV$, for example, $DP=D^+\pi^-,
D^0\pi^0,D^0\eta_8,D_s^+K^-$, can explain the enhancement of not
only $D^0\pi^0$ but also $D^0\eta$ via inelastic rescattering from
the class-I mode $\ov B^0\to D^+\pi^-$.} More precisely, elastic
FSIs will enhance the decay rate of $D^0\pi^0$ by a factor of
about 3 and suppress $D^+\pi^-$ slightly.

It has been realized that the isospin analysis proves to be useful
only if a few channels are open as the case of two-body
nonleptonic decays of kaons and hyperons. The isospin phases there
(or decay amplitude phases) are related to strong-interaction
eigenphases (for a recent discussion, see \cite{Suzuki}). For
example, one can identify the isospin phase shift in $K\to \pi\pi$
with the measured $\pi\pi$ strong-interaction phase at the energy
$\sqrt{s}=m_K^2$. However, when there are many channels open and
some channels coupled, as in $D$ and especially $B$ decays, the
decay phase is no longer the same as the eigenphase in the
$S$-matrix. Indeed, the $S$-matrix in general contains a parameter
describing inelasticity. Consider the decay $\ov B^0\to D^+\pi^-$
as an example. The state $D^+\pi^-$ couples to not only
$D^0\pi^0$, but also $D^0\eta,\,D^0\eta'$, $D\pi\pi\pi$ channels,
$\cdots$, etc. It has been argued that in the heavy quark limit
the $B$ decay is dominated by multiparticle inelastic rescattering
\cite{Donoghue}. As a consequence, even if elastic $D^{(*)}\pi$
scattering is measured at energies $\sqrt{s}=m_B$, the isospin
phases appearing in (\ref{isospin}) or (\ref{isospinfsi}) cannot
be identified with the measured strong phases. Moreover, the
isospin amplitudes are not conserved by inelastic FSIs. Therefore,
the isospin analysis presented before should be regarded as an
intermediate step for describing physical decay amplitudes.

Nevertheless, the isospin decomposition of $\ov B\to D\pi$
amplitudes in Eq. (\ref{isospin}) or (\ref{isospinfsi}) is still
valid. The isospin analysis is useful in some aspects. First, it
provides an independent check on the relative phases among three
decay amplitudes. Second, the deviation of
$|A_{1/2}/(\sqrt{2}A_{3/2})|$ from unity measures the degree of
departure from the heavy quark limit \cite{NP}. Third, the
deviation of $a_2$ from $a_2^{\rm eff}$ characterizes the
importance of (soft) FSI contributions to the color-suppressed
quark diagram, recalling that $a_{1,2}^{\rm eff}$ are defined for
the topologies without FSIs. This point will be elucidated more
below.

As stressed in \cite{CC}, the topological quark graphs are meant
to have all strong interactions included. Hence, they are {\it
not} Feynman graphs. For example, the genuine $W$-exchange
topology in $\ov B\to D\pi$ decay consists of not only the
short-distance $W$-exchange diagram but also the rescattering
graph in which $\ov B^0\to D^+\pi^-$ is followed by the strong
interaction process: $(D^+\pi^-)_{I=1/2}\to$ scalar resonances
$\to D^0\pi^0$. Likewise, the process with inelastic rescattering
from the leading $\T$ amplitude into $D^0\pi^0$ via quark exchange
has the same topology as the color-suppressed tree diagram $\C$
\cite{Neubert}. Therefore, color-suppressed tree and $W$-exchange
topologies receive short-distance and long-distance contributions.

From Tables III and V we see that $R_{D\pi}=|a_2(D\pi)/a_2^{\rm
eff}(D\pi)|\approx 1.75$ and $R_{D^*\pi}=|a_2(D^*\pi)/a_2^{\rm
eff}(D^*\pi)|\approx 1.95$. The corresponding quantities in
$D\to\ov K\pi$ decays are $R_{\ov K\pi}\approx 2.0$ and $R_{\ov
K^*\pi}\approx 1.7$, respectively. Therefore, although the
relative phase $59^\circ$ ($63^\circ$) between $B^0\to
D^{0(*)}\pi^0$ and $B^0\to D^{+(*)}\pi^-$ is significantly reduced
from the phase $150^\circ$ between $D^0\to\ov K^{0(*)}\pi^0$ and
$D^0\to K^{-(*)}\pi^+$ \cite{Rosner}, the ratio $R$ does not
decrease sizably from charm to bottom and, in contrast, it
increases for the $VP$ case.  It is thus anticipated that in both
$D\to \ov K\pi$ and $\ov B\to D\pi$ decays, the soft FSI
contributions to the color-suppressed topology $\C$ are dominated
by inelastic rescattering \cite{Donoghue}.\footnote{The quark
diagram $W$-exchange in $D\to \ov PP$ decays and its phase
relative to the topological amplitude $\T$ are dominated by nearby
resonances in the charm mass region \cite{Zen}, as shown
explicitly in \cite{CT}.} Since $\eta$ and $\omega$ are isospin
singlets, the conventional isospin analysis of FSIs is no longer
applicable to the final states involving $\eta$ or $\omega$. The
fact that the predicted $\ov B^0\to D^{(*)0}\eta$ rates based on
the assumption $a_2(D^{(*)}\eta)=a_2(D^{(*)}\pi)$ are consistent
with experiment (see Table IV) supports the notion that FSIs in
$B$ decay are indeed highly inelastic.

\section{Discussion and Conclusion}
Beyond the phenomenological level, it is desirable to have a
theoretical estimate of $a_2(D\pi)$. Unfortunately, contrary to
the parameter $a_1(D\pi)$, $a_2(D\pi)$ is not calculable in the
QCD factorization approach owing to the presence of infrared
divergence caused by the gluon exchange between the emitted $D^0$
meson and the $(\ov B^0\pi^0)$ system. In other words, the
nonfactorizable contribution to $a_2$ is dominated by
nonperturbative effects. Nevertheless, a rough estimate of $a_2$
by treating the charmed meson as a light meson while keeping its
highly asymmetric distribution amplitude yields $a_2(D\pi)\approx
0.25\exp(-i40^\circ)$ \cite{BBNS}. Evidently, large power
corrections from long-distance FSI effects are needed to account
for the discrepancy between theory and experiment for $a_2(D\pi)$.
The rescattering contribution via quark exchange, $D^+\pi^-\to
D^0\pi^0$, to the topology $\C$ in $\ov B^0\to D^0\pi^0$ has been
estimated in \cite{Blok} using $\rho$ trajectory Regge exchange.
It was found that the additional contribution to $D^0\pi^0$ from
rescattering is mainly imaginary: $a_2(D\pi)/a_2(D\pi)_{\rm
without~FSIs}=1+0.61{\rm exp}(73^\circ)$. This analysis suggests
that the rescattering amplitude can bring a large phase to
$a_2(D\pi)$ as expected.

In QCD factorization, $a_2(\pi\pi)$ or $a_2(K\pi)$ is found to be
of order 0.20 with a small strong phase (see e.g. \cite{BBNS1}).
The fact that the magnitude of $a_2(D\pi)$ is larger than the
short-distance one, $a_2(K\pi)$ or $a_2^{\rm eff}(D\pi)$, should
not be surprising because the former includes all possible FSIs,
while the latter is defined without long-distance FSIs. In other
words, $a_2(D\pi)$ include many possible long-distance effects. In
the language of isospin analysis, we see from Eq. (\ref{ampfsi})
that
 \be
 a_2(D\pi)=a_2^{\rm eff}(D\pi)-{2ha_1^{\rm eff}(D\pi)-a_2^{\rm eff}
 (D\pi)\over
 3}\left(e^{i(\delta_{1/2}-\delta_{3/2})}-1\right),
 \en
where we have neglected $W$-exchange and
 \be
 h={f_\pi\over f_D}\,{m_B^2-m_D^2\over
 m_B^2-m_\pi^2}\,{F_0^{BD}(m_\pi^2)\over
 F_0^{B\pi}(m_D^2)}.
 \en
It follows from Eq. (\ref{A13}) and Table V that
$a_2(D\pi)/a_2^{\rm eff}(D\pi)\approx 1.65\,{\rm exp}(56^\circ)$.
It is worth remarking that $a_2(\J K)$ in $B\to \J K$ decay is
calculable in QCD factorization; the theoretical result $|a_2(\J
K)|=0.19^{+0.14}_{-0.12}$ \cite{CYJpsiK} is consistent with the
data $0.26\pm 0.02$ \cite{a1a2}. Hence it remains to understand
why $|a_2(D\pi)|$ is larger than $|a_2(D^*\pi)|$ and $|a_2(\J K)|$
or why (soft) final-state interaction effects are more important
in $D\pi$, $D^*\pi$ than in $\J K$ final states.

To conclude, the recent measurements of the color-suppressed modes
$\ov B^0\to D^{(*)0}\pi^0$ imply non-vanishing relative FSI phases
among various $\ov B\to D\pi$ decay amplitudes. Depending on
whether or not FSIs are implemented in the topological
quark-diagram amplitudes, two solutions for the parameters $a_1$
and $a_2$ are extracted from data using various form-factor
models. It is found that $a_2$ is not universal: $|a_2(D\pi)|\sim
0.40-0.55$ and $|a_2(D^*\pi)|\sim 0.30-0.45$ with a relative phase
of order $60^\circ$ between $a_1$ and $a_2$. If FSIs are not
included in quark-diagram amplitudes from the outset, we have
$a_2^{\rm eff}(D\pi)\sim 0.23-0.32,~a_2^{\rm eff} (D^*\pi)\sim
0.16-0.22\,.$ The large value of $|a_2(D\pi)|$ compared to
$a_2^{\rm eff}(D\pi)$ or naive expectation implies the importance
of long-distance FSI contributions to color-suppressed internal
$W$-emission via final-state rescatterings of the color-allowed
tree amplitude.

\vskip 2 cm  \acknowledgments We would like to thank Hsiang-nan
Li, Alexey A. Petrov, Zhi-zhong Xing and Kwei-Chou Yang for
delighting discussions. We also wish to thank Physics Department,
Brookhaven National Laboratory for its hospitality. This work was
supported in part by the National Science Council of R.O.C. under
Grant No. NSC90-2112-M-001-047.


\end{document}